\documentclass[pra,twocolumn,showpacs,preprintnumbers,amsmath,amssymb]{revtex4}
\usepackage{graphicx}% Include figure files
\usepackage{dcolumn}% Align table columns on decimal point
\usepackage{bm}% bold math
\usepackage{amsthm}
\usepackage{amscd}
\usepackage{epsfig}
\usepackage{color}
\input xy
\xyoption{all}

\newcommand{\ket}[1]{\left| #1 \right\rangle}
\newcommand{\bra}[1]{\left\langle #1 \right |}
\newtheorem{theorem}{Theorem}

\DeclareMathOperator{\tr}{tr}

\DeclareMathOperator{\Sym}{Sym}

\newcommand{\R}{{\mathbb R}}

\newcommand{\C}{{\mathbb C}}

\newcommand{\Id}{{\rm Id}}

\begin{document}

\title{Entanglement classes of symmetric Werner states}

\author{David W. Lyons}
  \email{lyons@lvc.edu}
\author{Scott N. Walck}
  \email{walck@lvc.edu}
\affiliation{Lebanon Valley College, Annville, PA 17003}

\date{revised 14 September 2011}

\begin{abstract}
The symmetric Werner states for $n$ qubits, important in the study of
quantum nonlocality and useful for applications in quantum information,
have a surprisingly simple and elegant structure in terms of tensor
products of Pauli matrices.  Further, each of these states forms a
unique local unitary equivalence class, that is, no two of these states
are interconvertible by local unitary operations.
\end{abstract}

\pacs{03.67.Mn}

\maketitle

% \section{Introduction}

Quantum information, motivated by practical applications in computation
and cryptography, has been instrumental in shedding light on fundamental
theoretical questions in physics and computer science. These include
violation of Bell inequalities and local hidden variable
theories~\cite{bell64,werner89,greenberger89}, new
proofs of classical information theorems~\cite{druckerdewolf2011}, and
new physical principles such as information
causality~\cite{pawloskietalinfcausality2009}.
%% Quantum information has provided insights on classical open problems in
%% complexity theory [cite something]. 
Basic questions about states have been a driving theme. For example,
when is a given state separable? When does it exhibit correlations that
are non-classical? Considering quantum states as resources raises the
question of interconvertibility. When can one given state be transformed
by local operations into a second given state? In general, these
questions are difficult. In this article, we consider a class of states
that has been demonstrated to have interesting and useful properties for
which we give a structure theorem and answer the interconvertibility
question. We introduce a novel analysis based on the identification of
symmetric density matrices with real polynomials in three variables and
apply the representation theory of $SO(3)$ on this space of polynomials.

The symmetric Werner states lie in the intersection of two important
classes of composite states of subsystems of equal dimension.  Symmetric
states, that is, states that are invariant under permutation of
subsystems, are the subject of recent work including: geometric measure
of entanglement
\cite{aulbach2010,aulbach2010b,markham2010,PhysRevA.82.032301,1367-2630-12-8-083002},
efficient tomography \cite{toth2010}, classification of states
equivalent under stochastic local operations and classical communication
(SLOCC) \cite{bastin2009,bastin2010,PhysRevLett.105.200501}, and our own
work on classification of states equivalent under local unitary (LU)
transformations \cite{symmstatespaper,symmstates2}.  Werner states,
defined to be those states invariant under the action of any particular
single qubit unitary operator acting on all $n$ qubits, have found a
multitude of uses in quantum information science.  Originally introduced
in 1989 for two particles~\cite{werner89} to distinguish between
classical correlation and Bell inequality satisfaction, Werner states
have found use in the description of noisy quantum
channels~\cite{lee00}, as examples in nonadditivity
claims~\cite{shor01}, and in the study of deterministic
purification~\cite{short09}.  In what may prove to be a practical
application to computing in noisy environments, Werner states comprise
decoherence-free subspaces for collective
decoherence~\cite{zanardi97,lidar98,bourennane04}.

An understanding of structure and entanglement properties of mixed
Werner states is known for bipartite and tripartite systems of arbitrary
dimension~\cite{werner89,eggelingwernerPhysRevA.63.042111}, but remains
an open problem for higher numbers of component subsystems. It is
natural to restrict the problem to subclasses that might be more
tractable. In previous work~\cite{su2blockstates} we have fully
classified local unitary equivalence classes of {\em pure} Werner
states. In this article, we consider the case of {\em symmetric} Werner
states (pure and mixed), which include the singlet and the uniform
mixture of symmetric Dicke states.  The symmetric Werner states for $n$
qubits have a surprisingly simple and elegant structure: their density
matrices consist of linear combinations of symmetrized products of
$\sigma_x \otimes \sigma_x + \sigma_y \otimes \sigma_y + \sigma_z
\otimes \sigma_z$ and the identity. Further, each of these states forms
a unique local unitary equivalence class, that is, no two of these
states are interconvertible by local unitary operations.

We show how symmetric mixed states of $n$ qubits can be
represented as real polynomials in three variables.  We then show how to
use representation theory of $SO(3)$ on polynomials to obtain the main
result, Theorem~\ref{symm-wern-stat} below.

\section{Preliminaries}\label{prelimsection}

Let ${\cal W}$ denote the 4-dimensional real
vector space of $2 \times 2$ Hermitian matrices.  A convenient basis for
${\cal W}$ is $\{ \sigma_0, \sigma_1, \sigma_2, \sigma_3 \}$, where $\sigma_0$
is the $2 \times 2$ identity matrix, and $\sigma_1=\sigma_x$,
$\sigma_2=\sigma_y$, and $\sigma_3=\sigma_z$ are the Pauli matrices.

The set of $n$-qubit density matrices is a proper subset of the vector space ${\cal
  W}^{\otimes n}$, where every element $\rho$ (whether or not $\rho$ is
positive or has trace 1) can be uniquely written in the form $\rho=\sum_I
s_I\sigma_I$, where $I=i_1i_2\ldots i_n$ is a multiindex with
$i_k=0,1,2,3$ for $1\leq k\leq n$, and $\sigma_I$ denotes
$$\sigma_I = \sigma_{i_1}\otimes \sigma_{i_2}\otimes \cdots \otimes
\sigma_{i_n},$$
with real coefficients $s_I$.

Given a permutation $\pi$ of $\{1,2,\ldots,n\}$, let $P_\pi$ denote the operator
on Hilbert space that carries out the corresponding permutation of qubits.
Here are two examples.
\begin{align*}
P_{(23)} \ket{11000} &= \ket{10100} \\
P_{(23)} \left( \sigma_j \otimes \sigma_k \otimes \sigma_l \right) P_{(23)}^{-1}
 &= \sigma_j \otimes \sigma_l \otimes \sigma_k
\end{align*}
Define a symmetrization operator on density matrices as
\[
\Sym(\rho) = \frac{1}{n!} \sum_{\pi} P_\pi \rho P_\pi^{-1} .
\]
An $n$-qubit symmetric density matrix $\rho$
is one for which $P_\pi \rho P_\pi^{-1} = \rho$ for all permutations $\pi$,
or equivalently, one for which $\Sym(\rho) = \rho$.
We denote by $\Sym^n {\cal W}$ the $n$-fold symmetric power of ${\cal W}$.  It is the subspace
of elements of ${\cal W}^{\otimes n}$ that are invariant under qubit permutation.
Every $n$-qubit symmetric density matrix $\rho$ is an element
of $\Sym^n {\cal W}$, and can be written
\begin{equation}\label{symrho}
\rho = \frac{1}{2^n} \sum c_{n_1n_2n_3} \Sym \left(
  \sigma_0^{\otimes n_0} \otimes \sigma_1^{\otimes n_1} \otimes
  \sigma_2^{\otimes n_2} \otimes \sigma_3^{\otimes n_3} \right) ,
\end{equation}
where the sum is over non-negative integers $n_0$, $n_1$, $n_2$, and $n_3$
such that $n_0 + n_1 + n_2 + n_3 = n$.
The coefficients $c_{n_1n_2n_3}$ are real.
The collection of $n$-qubit symmetric density matrices is a proper subset of
$\Sym^n {\cal W}$, since the latter contains
Hermitian matrices that are not positive semi-definite, and Hermitian matrices for
which the trace is not 1.

Let $\R_n[x,y,z]$ be the set of polynomials of degree at most $n$
in three variables $x$, $y$, and $z$ with
real coefficients.
For each $n$, there is a linear map $F_n: \Sym^n {\cal W} \to \R_n[x,y,z]$ defined by
\[
\frac{1}{2^n} \Sym \left(
  \sigma_0^{\otimes n_0} \otimes \sigma_1^{\otimes n_1} \otimes
  \sigma_2^{\otimes n_2} \otimes \sigma_3^{\otimes n_3} \right) \mapsto
x^{n_1} y^{n_2} z^{n_3} .
\]
In this way, we may associate a polynomial of degree at most $n$ with each
$n$-qubit symmetric mixed state.  The polynomial associated with
(\ref{symrho}) is
\begin{equation}
F_n(\rho) = \sum_{\stackrel{n_1,n_2,n_3}{n_1+n_2+n_3 \leq n}} c_{n_1n_2n_3} x^{n_1} y^{n_2} z^{n_3} .
\end{equation}
For each $n$, the map $F_n$ is an invertible linear map.
Table \ref{exampletable} lists these polynomials for some example
symmetric mixed states.

\begin{table}
\begin{center}
\begin{tabular}{lrc}
Symmetric Mixed State $\rho$ & $n$ & $F_n(\rho)$ \\ \hline
 $\ket{0} \bra{0}$                                     & 1 & $1 + z$ \\
 $\ket{1} \bra{1}$                                     & 1 & $1 - z$ \\
$\ket{+} \bra{+} = \frac{1}{2} (\ket{0} + \ket{1}) (\bra{0} + \bra{1})$ & 1 & $1 + x$ \\
Totally mixed, $\frac{1}{2} \ket{0} \bra{0} + \frac{1}{2} \ket{1} \bra{1}$ & 1 & $1$ \\
 $\ket{00} \bra{00}$                                    & 2 & $(1 + z)^2$ \\
Totally mixed                                           & 2 & $1$ \\
 $\frac{1}{2} \ket{00} \bra{00} + \frac{1}{2} \ket{11} \bra{11}$ & 2 & $1 + z^2$ \\
 $\frac{1}{2} \ket{0} \bra{0} \otimes \frac{I}{2}
  + \frac{1}{2} \frac{I}{2} \otimes \ket{0} \bra{0}$ & 2 & $1 + z$ \\
Singlet, $\frac{1}{2} (\ket{01} - \ket{10})(\bra{01} - \bra{10})$ & 2 & $1 - x^2 - y^2 - z^2$ \\
  $\frac{1}{2} (\ket{00} - \ket{11})(\bra{00} - \bra{11})$ & 2 & $1 - x^2 + y^2 + z^2$ \\
  $\frac{1}{2} (\ket{00} + \ket{11})(\bra{00} + \bra{11})$ & 2 & $1 + x^2 - y^2 + z^2$ \\
  $\frac{1}{2} (\ket{01} + \ket{10})(\bra{01} + \bra{10})$ & 2 & $1 + x^2 + y^2 - z^2$ \\
Uniform Dicke mixture  & 2 & $1 + \frac{1}{3} (x^2 + y^2 + z^2)$ \\
  $\ket{000} \bra{000}$                                    & 3 & $(1 + z)^3$ \\
GHZ, $\frac{1}{2} (\ket{000} + \ket{111})(\bra{000} + \bra{111})$ & 3 & $1 + 3 z^2 + x^3 - 3 x y^2$ \\
W          & 3 & $(1 + z) \cdot {}$ \\
           &   & $(1 + 2x^2 + 2y^2 - z^2)$ \\
Uniform Dicke mixture  & 3 & $1 + x^2 + y^2 + z^2$ \\
Uniform Dicke mixture  & 4 & $1 + 2(x^2 + y^2 + z^2)$ \\
                       &   & ${} + \frac{1}{5}(x^2 + y^2 + z^2)^2$ \\
%%  $\frac{1}{3} \ket{D_2^{(0)}} \bra{D_2^{(0)}}
%% + \frac{1}{3} \ket{D_2^{(1)}} \bra{D_2^{(1)}}
%% + \frac{1}{3} \ket{D_2^{(2)}} \bra{D_2^{(2)}}$
\end{tabular}
\end{center}
\caption{Polynomials for some symmetric mixed states.
The density matrix for the $W$ state is
$\rho_W = \frac{1}{3} (\ket{100} + \ket{010} + \ket{001})
                      (\bra{100} + \bra{010} + \bra{001})$.
The totally mixed 1-qubit state is
$\frac{I}{2} = \frac{1}{2} \ket{0} \bra{0} + \frac{1}{2} \ket{1} \bra{1}$.}
\label{exampletable}
\end{table}

Since $F_n$ is a linear map, the polynomial for a mixture of symmetric mixed states
is the mixture of the polynomials.
\[
F_n(p_1 \rho_1 + p_2 \rho_2) = p_1 F_n(\rho_1) + p_2 F_n(\rho_2)
\]

A product of polynomials $\R_n[x,y,z] \times \R_m[x,y,z] \to \R_{n+m}[x,y,z]$
represents the symmetrized tensor product of states.
\[
F_n(\rho_1) F_m(\rho_2) = F_{n+m}(\Sym(\rho_1 \otimes \rho_2))
\]

Let $g \in SU(2)$.  Define $T_g: \Sym^n {\cal W} \to \Sym^n {\cal W}$ to
be the symmetric transformation of each qubit by $g$.
\[
T_g(\rho) = g^{\otimes n} \rho (g^{\dagger})^{\otimes n}
\]
If we define $R_g: \R_n[x,y,z] \to \R_n[x,y,z]$ to be the transformation on polynomials
defined by
\begin{equation}\label{su2actiononpolys}
R_g(f)(x,y,z) = f \left( (x,y,z) \Phi(g) \right) ,  
\end{equation}
where $f$ is a polynomial, $\Phi: SU(2) \to SO(3)$ is the
homomorphism~\footnote{$\Phi$ is given in a natural way by the adjoint
  action $SU(2)\to SO(su(2))$, so that $\Phi(g)(M) = gMg^\dagger$, and we
  identify $su(2)$ with $\R^3$ by $A\leftrightarrow (1,0,0)$,
  $B\leftrightarrow (0,1,0)$, $C\leftrightarrow
  (0,0,1)$. See~\cite{b-td:rclg}.  } that associates a rotation in
$\R^3$ with each $2 \times 2$ unitary, and $(x,y,z) \Phi(g)$ denotes a
row vector multiplied by a $3 \times 3$ orthogonal matrix, then the
following diagram commutes.
\[
\begin{CD}
\R_n[x,y,z] @>R_g>> \R_n[x,y,z] \\
  @AF_nAA           @AF_nAA \\
\Sym^n {\cal W}  @>T_g>> \Sym^n {\cal W}  \\
\end{CD}
\]

It is curious in Table \ref{exampletable} that the pure $W$ state appears
as the product of other polynomials, and hence as the
symmetrized tensor product of Hermitian matrices.
Since a symmetrization is a mixture, it seems contradictory
that the pure $W$ state could be a mixture.
The resolution is that the factor $1 + 2x^2 + 2y^2 - z^2$
corresponds to a Hermitian matrix with a negative eigenvalue, and
hence does not represent a density matrix.
It is an interesting question to ask whether there
is physical significance in the factor-ability of the $W$ state.

\section{Symmetric Werner states}

\begin{theorem}\label{symm-wern-stat}
Let $\rho$ be an $n$-qubit symmetric mixed state.
We claim that $g^{\otimes n} \rho
(g^{\dagger})^{\otimes n} = \rho$ for all $g \in SU(2)$ if and only if
$F_n(\rho)$ is a linear combination of terms of the form $(x^2 + y^2 +
z^2)^m$, for $0 \leq m \leq n/2$. Further, any two such states are local
unitarily inequivalent. That is, there is no product $g=g_1\otimes
g_2\otimes \cdots \otimes g_n$ of local unitaries $g_i$ such that
$g\rho g^\dagger = \rho'$ unless $\rho =\rho'$.
\end{theorem}

It is clear that polynomials of the form
$F_n(\rho) = \sum_{m=0}^{\lfloor n/2 \rfloor} b_m (x^2 + y^2 + z^2)^m$
are invariant under the $SU(2)$ action~(\ref{su2actiononpolys}) since
$x^2 + y^2 + z^2$ is invariant under any rotation in $SO(3)$.

Conversely, suppose that $g^{\otimes n} \rho (g^{\dagger})^{\otimes n} =
\rho$ for all $g \in SU(2)$.  Then $R_g(F_n(\rho)) = F_n(\rho)$ for all
$g \in SU(2)$.  We seek to find the subspace of $\R_n[x,y,z]$ that
transforms as the trivial irreducible representation of $SO(3)$.
Homogeneous polynomials in three variables $x$, $y$, and $z$ are known
to be reducible representations of $SO(3)$ \cite{sternberg}\footnote{To
  be precise, the cited work considers the representation
  $\C[x,y,z]$. In the case of $SO(3)$, the irreducible submodules of
  this complex representation are in one-to-one correspondence with the
  {\em real} irreducible submodules of $\R[x,y,z]$ via
  complexification. See~\cite[Ch.2 Sec.6]{b-td:rclg}.}.  If $V_l$ is the
irreducible representation of $SO(3)$ with dimension $2l+1$, then the
homogeneous polynomials of degree $p$ in three variables decompose into
irreducible representations as
\begin{equation}
\bigoplus_{j=0}^{\lfloor p/2 \rfloor} V_{p-2j} .
\label{homopolyirred}
\end{equation}
When expressed as a sum of irreducible
representations, the homogeneous polynomials of degree $p$
in three variables contain one dimension of the trivial
representation ($V_0$) if $p$ is even, and zero dimensions
if $p$ is odd \cite{sternberg}.  The vector space $\R_n[x,y,z]$ of polynomials
of degree at most $n$ is a direct sum of vector spaces of
homogeneous polynomials of degree $p$, for $0 \leq p \leq n$.
Therefore, the space $\R_n[x,y,z]$ contains $\lfloor n/2 \rfloor + 1$
dimensions of the trivial representation.
Since we have identified all of these dimensions, $F_n(\rho)$
must be a linear combination of the polynomials given.

Finally, let $\rho,\rho'$ be symmetric Werner states, and suppose there
is a local unitary operation $g=g_1\otimes g_2\otimes \cdots \otimes
g_n$ such that $g\rho g^\dagger = \rho'$. We show in~\cite{symmstates2}
that for $n\geq 3$ there exists an element $h\in SU(2)$ such that $h^{\otimes n} \rho
(h^\dagger)^{\otimes n} = g\rho g^\dagger$. Because $\rho$ is a Werner
state, we have $\rho=\rho'$. For $n=2$, it is straightforward, albeit tedious, to check
that the coefficients of $\lambda, \lambda^2$ in the characteristic
polynomial $\det(\rho(a)-\lambda \Id)$ for the state
$\rho(a):=\Id/4+\frac{a}{4}(\sigma_x\otimes \sigma_x +\sigma_y\otimes \sigma_y
+\sigma_z\otimes \sigma_z)$ are $\frac{1-3a^2-2a^3}{16}$ and
$\frac{3-3a^2}{8}$, respectively. This implies that if $\rho(a)$ is
          local unitary equivalent to $\rho(a')$, then $a=a'$.

\section{Summary and Outlook}

We have shown how symmetric Werner states have a simple structure
expressed in a basis of tensors of Pauli matrices, and that
decompositions in this basis are local unitary invariants. Natural next
questions are
\begin{itemize}
\item Can we give bounds on coefficients in the polynomials $F_n(\rho)$?
\item Can we find conditions for separability using these coefficients?
\item Can we say what level of reduced density matrices determines
  symmetric Werner states? (That is, if we are given a collection of $k$-qubit reduced
  density matrices for some $k$, can we determine whether there is a unique $\rho$
  that has these reduced density matrices?)
\end{itemize}
Evidence that the last question is not just wishful thinking is that the
partial trace operation is particularly nice in Pauli tensor
coordinates. We have
$$\tr_k (\sigma_{i_1}\otimes\cdots \otimes \sigma_{i_n}) = 
\left\{
\begin{array}{cc}
 2 \sigma_{i_1}\otimes\cdots \widehat{\sigma_{i_k}} \cdots  \otimes
  \sigma_{i_n} & \mbox{if $i_k=0$}\\
0 & \mbox{otherwise}
\end{array}
\right.
$$ where the wide hat symbol means omit the $k$th factor.  Progress on
any such aspect of the $N$-representability problem would be of
interest.

\medskip

{\em Acknowledgments.}  
This work has been supported by National Science Foundation grant
\#PHY-0903690. D.L. thanks Geza Toth for helpful discussions.

%%% To make a version that includes the bibliography in a single file:
%%% 1) latex filename.tex
%%% 2) cp filename.tex filenamebib.tex
%%% 3) edit filenamebib.tex to include filename.bbl after this comment
%%% 4) edit filename.tex to comment out \bibliography{bibfilename}

% \bibliography{qimain}

\end{document}